\documentclass[12pt,preprint2]{aastex}
\usepackage{emulateapj5}
\usepackage{onecolfloat5}

\usepackage{natbib}

\newcommand{\rxj}{RX~J1308.6+2127}
\newcommand{\rxjw}{RX~J1856.5$-$3754}
\newcommand{\chandra}{\textit{Chandra}}

\newcommand{\hst}{\textit{HST}}

\newcommand{\rxjk}{RX~J0720.4$-$3125}
\newcommand{\rosat}{\textit{ROSAT}}
\newcommand{\lfxo}{\ensuremath{\log(f_{\rm X}/f_{\rm opt})}}
\newcommand{\til}{\raisebox{-0.7ex}{\ensuremath{\tilde{\;}}}}
\newcommand{\gsim}{\gtrsim}
\newcommand{\expnt}[2]{\ensuremath{#1 \times 10^{#2}}}   % scientific notation

\slugcomment{Accepted to ApJL}
\begin{document}

\shorttitle{A Probable Counterpart for \rxj}
\shortauthors{Kaplan et al.}

\twocolumn[
\title{A Probable Optical Counterpart for the Isolated Neutron Star \rxj}
\author{D.~L.~Kaplan, S.~R.~Kulkarni}
\affil{Department of Astronomy, 105-24 California Institute of
Technology, Pasadena, CA 91125, USA}
\email{dlk@astro.caltech.edu, srk@astro.caltech.edu}
\and
\author{M.~H.~van~Kerkwijk}
\affil{Sterrenkundig Instituut, Universiteit Utrecht,
Postbus 80000, 3508 TA Utrecht, The Netherlands} 
\email{M.H.vanKerkwijk@astro.uu.nl}

\begin{abstract}
Using a very deep observation with \hst/STIS, we have searched for an
optical counterpart to the nearby radio-quiet isolated neutron star
\rxj\ (RBS~1223).  We have identified a single object in the 90\%
\chandra\ error circle that we believe to be the optical counterpart.
This object has $m_{\rm 50CCD}=28.56\pm0.13$~mag, which translates
approximately to an unabsorbed flux of $F_{\lambda}=\expnt{(1.7 \pm
0.3)}{-20}\mbox{ ergs s}^{-1}\mbox{ cm}^{-2}\mbox{ \AA}^{-1}$ at
5150~\AA\ or an X-ray-to-optical flux ratio of $\lfxo=4.9$.  This flux
is a factor of $\approx 5$ above the extrapolation of the black-body
fit to the X-ray spectrum, consistent with the optical spectra of
other isolated neutron stars.  Without color information we cannot
conclude that this source is indeed the counterpart of \rxj.  If not,
then the counterpart must have $m_{\rm 50CCD} > 29.6$~mag,
corresponding to a flux that is barely consistent with the
extrapolation of the black-body fit to the X-ray spectrum.

\end{abstract}

]

\section{Introduction}
\label{sec:intro}
Neutron stars have been regarded as natural laboratories
for matter denser than can be obtained by heavy-ion
accelerators.  The basic physics is summarized by the mass and
radius, with larger radii for a given mass favoring 
stiffer equations-of-state (EOS; \citealt{lp00}). It is against this backdrop
that one recognizes that one of the   
major outcomes of the all-sky survey undertaken by 
the X-ray satellite \textit{ROSAT} was the systematic identification
of the nearest neutron stars (see reviews by \citealt{motch01} and \citealt{ttzc00}).

\rxj\ (also known as RBS~1223) was identified as a candidate
neutron star from the \textit{ROSAT}
Bright Survey by \citet{shs+99} on the basis of its soft X-ray
spectrum (blackbody with $kT\approx 100$~eV), constant X-ray flux, and
lack of  optical counterpart. It now joins six other similar objects
(RX~J1856.5$-$3754, RX~J0720.4$-$3125, RX~J1605.3+3249, RX~J2143.0
+0654, RX~J0806.4$-$4123, and RX~J0420.0$-$5022;
\citealt*{wwn96}; \citealt{hmb+97,mhz+99}; \citealt*{hpm99}; \citealt{zct+01}) and three previously known
pulsars (Geminga, PSR~B0656+14, and PSR~B1055$-$52) in the sample of
nearby $10^{6}$-yr neutron stars detected by the \rosat\ Bright Survey.

Of these objects, the five brightest (in terms of soft X-ray
count-rate)  have been well studied. PSR~0656+14 and PSR~B1055$-$52 are well
known radio pulsars, not particularly remarkable in any other way.
Geminga, first identified via is $\gamma$-ray emission (and thereby
dramatically demonstrating that radio pulsars can lose a large
fraction of their energy via $\gamma$-rays) is now generally considered
to be an ordinary pulsar whose radio beam we happen to miss.

In contrast, \rxjw\ and \rxjk\ are mysterious. Both sources have (as
expected) faint, blue, optical counterparts \citep{wm97,kvk98}, with
X-ray-to-optical flux ratios of $\lfxo\sim 5$.
\rxjw\ shows no significant pulsations \citep*{rgs02} and despite
significant investment of {\em Chandra} time, the X-ray spectrum is
featureless \citep{dmd+02}. There is no evidence for any non-thermal
emission \citep{vkk01}.  Conventional models for this source include a
weakly-magnetized cooling neutron star \citep{vkk01b} or an off-beam
radio pulsar \citep[like Geminga but without the $\gamma$-ray
emission;][]{br02}.  In contrast, \rxjk\ shows 8.4-s pulsations. It too 
exhibits a featureless X-ray spectrum (largely thermal;
\citealt{pmm+01}).  Again, conventional possibilities include an
off-beam radio pulsar but the long period would require that the
neutron star was born with either an unusually long period or an
unusually large magnetic field \citep{kkvkm02,zhc+02}.

Thus the five brightest (in soft X-rays) and presumably the
nearest neutron stars show a stunning diversity. Our
understanding of the nature of two (or perhaps even three) of these
sources is quite incomplete.

In this {\it Letter}, we  re-determine the position of \rxj\ from
archival \chandra\ analysis. We present radio observations of \rxj,
and we then discuss very deep optical
observations aimed at detecting its optical counterpart.
This source exhibits long-period pulsations with $P=5.16$~s.
However, unlike \rxjk\, a large period derivative has
been measured \citep{hhss02}.  If this is ascribed to magentic braking
then the implied dipole field strength is $B \gsim 10^{14}$~G, and
\rxj\ is a magnetar. \citet{kvk98} and \citet{hk98} advocated
the magnetar model for nearby long period pulsators because magnetars
have an additional source of heat (their magnetic fields) and thus are
warmer than ordinary neutron stars for a longer duration. 

\section{Observations \& Data Reduction}
\subsection{X-ray}
\label{sec:xray}
We used the 10-ks observation of \rxj\ from the \textit{Chandra X-ray
Observatory} described in \citet{hhss02} to determine the position of
the X-ray source.  The main change from the analysis presented by
\citet{hhss02} was that we corrected the spacecraft aspect by
$0\farcs23$ according to the CXC
prescription\footnote{\url{http://asc.harvard.edu/cal/ASPECT/fix\_offset/fix\_offset.cgi}}.
We measured the centroid of the X-ray source to be (J2000):
$\alpha=13^{\rm h}08^{\rm m}48\fs27$, $\delta=+21\degr27\arcmin
06\farcs78$, with statistical uncertainties of $\pm 0\farcs05$ in each
coordinate.  This position is consistent with the position determined
by \citet{hhss02} and with the \rosat\ HRI position, but it has been
tied (at least statistically) to the ICRS and is therefore preferable
for comparison with other data sets.  Overall, the
positions of X-ray sources that have been corrected for aspect errors
match optical positions (from the ICRS or Tycho) with a 90\%
confidence radius of $0\farcs6$, with a distribution that is highly
non-Gaussian\footnote{\url{http://asc.harvard.edu/cal/ASPECT/celmon/}}.

\subsection{Radio}

We observed \rxj\ with the Very Large Array (VLA) at 1.4~GHz on
2001 February 12 through 2001 February 14, with a total integration
time of 173~min.  Observations were done in the BnA configuration with
$2 \times50$-MHz bandwidths, giving a final beamsize of $1\farcs2
\times 3\farcs8$.  All data sets independently calibrated using {\tt
AIPS} then combined for imaging.  Imaging and self-calibration were
performed in {\tt difmap}.  The data were repeatedly cleaned and
self-calibrated (phase corrections only) until the solution converged.
After cleaning, we found rms map noise to be 0.032~mJy.

No emission from \rxj\ was found, giving a $3\sigma$ upper limit to
the flux of a point-source of 0.10~mJy.  This implies a radio
luminosity limit for \rxj\ of $0.05d_{700}^{2}\mbox{ mJy kpc}^{2}$
where the distance is $d=700 d_{700}$~pc \citep*{kvka02}.  Such a limit
is a factor of 10 below the luminosity PSR~J0205+6449 in 3C~58
\citep{csl+02} and a factor of 2 above the limit for Geminga
\citep{seiradakis92}.  In fact, it is below virtually all of the radio
pulsars younger than $10^{6}$~yr \citep{motch01}.

\begin{deluxetable}{c c c r c}
\tablecaption{Summary of Optical Observations\label{tab:obs}}
\tablewidth{0pt}
\tablehead{
\colhead{Telescope} & \colhead{Instrument} & \colhead{Date} &
\colhead{Exposure} & \colhead{Band} \\
 & & \colhead{(UT)} & \colhead{(sec)} & \\}
\startdata
Keck-II & ESI & 2000-May-03 & 3,500 & R \\
\hst & STIS & 2001-Jul-22 & 5,264 & 50CCD \\
\nodata & \nodata & 2001-Aug-04 & 15,880 & 50CCD \\
P200 & LFC & 2002-Mar-05 & 375 & $r$ \\
\enddata
\end{deluxetable}

\subsection{Optical}
\citet{shs+99} unsuccessfully searched for an optical counterpart to
\rxj, and determined limits of $B\gsim 26$~mag and $R\gsim 26$~mag
from observations at Keck.  The large X-ray-to-optical flux ratio
inferred from the ground based data strongly favored a neutron star
origin for \rxj.  Accordingly, we obtained data from the Space
Telescope Imaging Spectrometer (STIS) aboard \hst, listed in
Table~\ref{tab:obs}.  The data were taken with the CCD without a
filter (50CCD aperture), which gives an extremely broad spectral
response from 3000~\AA\ to 9000~\AA.  For these data, we drizzled
\citep{fh02} all the individual exposures onto a single image, for a
total exposure of 21,144~s (8 orbits).  We used a pixel scale of 0.5,
so that the final image had $0\farcs0254$ pixels.  For astrometric
purposes we also obtained data with the Echelle Spectrograph and
Imager (ESI; \citealt{sbe+02}) at the Keck-II telescope and the Large
Format Camera (LFC) at the Palomar 200-inch telescope.  We performed
standard reduction of the ground-based data in \texttt{IRAF},
subtracting bias images, flat-fielding, and stacking the exposures.
We referenced the astrometry to the latest version of the Guide Star
Catalog\footnote{\url{http://www-gsss.stsci.edu/support/data\_access.htm}}
(GSC~2.2).  After applying a distortion solution (M.~Hunt 2002,
personal
communication\footnote{\url{http://wopr.caltech.edu/{\til}mph/lfcred/}}),
we identified 142 unsaturated stars on the LFC image, solved for
rotation, zero-point, and plate scale (the same terms were used for
all subsequent astrometric solutions), and got residuals of $0\farcs4$
in each coordinate.

\begin{figure}[t]
\plotone{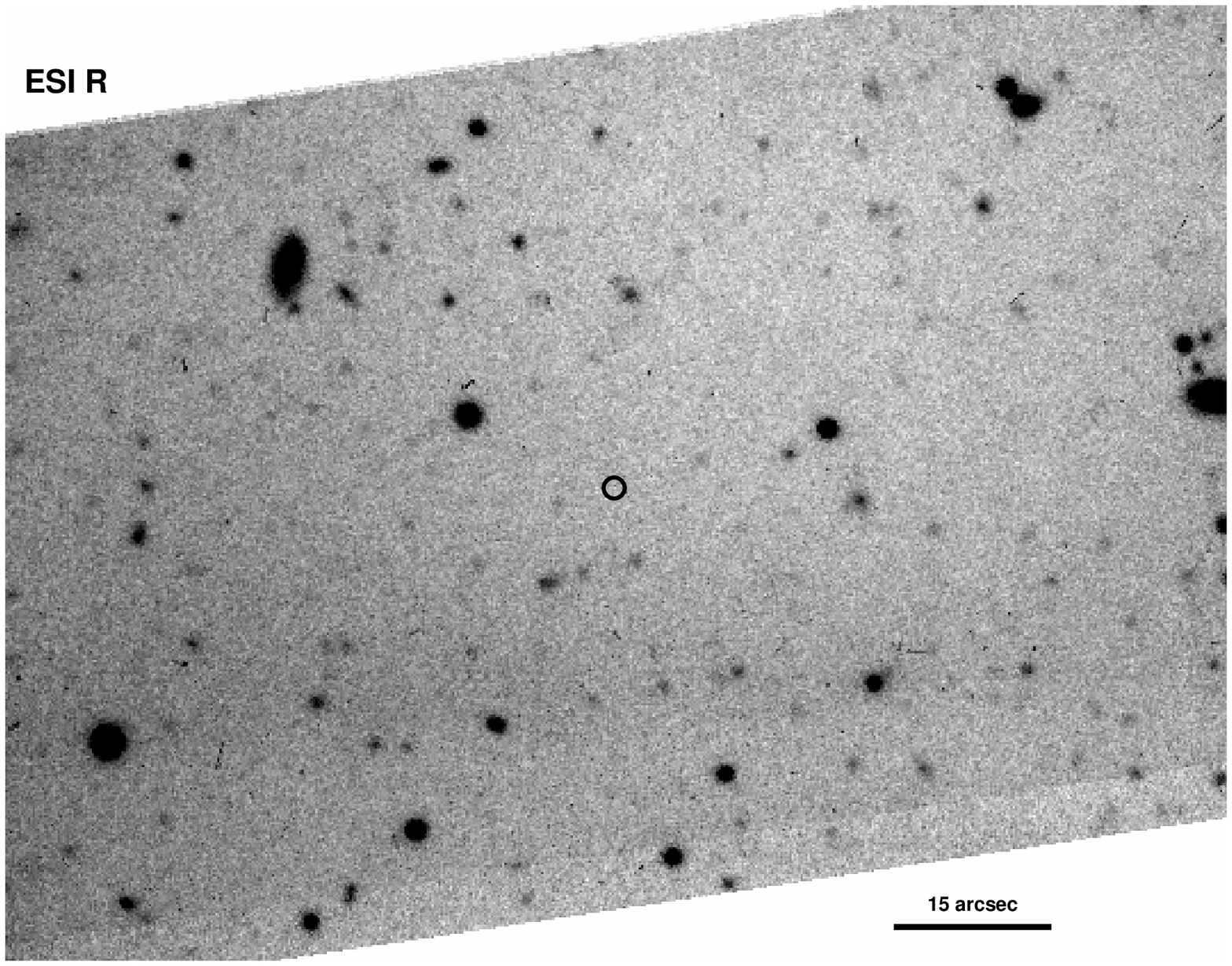}
\caption{Keck ESI image of the field around \rxj.  The image is
  $\approx 2\arcmin$ on a side, with North up and East to the left.
  The $1\farcs0$-radius \chandra\ error circle is shown.}
\label{fig:esi}
\end{figure}

We used 13 sources on the LFC image to transfer the astrometric
solution to the ESI image, with residuals of $0\farcs14$ in each
coordinate.  From the ESI image (Figure~\ref{fig:esi}), we identified 10 stars that we used
to go to the STIS image, and obtained residuals of $0\farcs06$ in each
coordinate.  Assuming a $0\farcs3$ intrinsic
uncertainty\footnote{\url{http://www-gsss.stsci.edu/gsc/gsc2/calibrations/astrometry/astrometry.htm\#method}}
for the GSC~2.2, we then have overall uncertainties of $0\farcs3$ in
each coordinate for the STIS image.  With the $0\farcs6$ 90\% radius
for the X-ray astrometry, we estimate a final 90\% confidence radius
of $\approx 1\farcs0$.

\section{Analysis \& Discussion}
In the following, we use the results of the spectroscopic fits of
\citet{hhss02} to the \chandra\ data.  Specifically, we take
$N_{H}=\expnt{(2.4\pm 1.1)}{20}\mbox{ cm}^{-2}$, $kT=91\pm 1$~eV and
$R=(6.5\pm 0.3) d_{700}$~km, where the normalization comes from the
\chandra\ count-rate. This spectrum implies an unabsorbed flux of
$\expnt{(3.5\pm 0.3)}{-21}\mbox{ ergs s}^{-1}\mbox{ cm}^{-2}\mbox{
\AA}^{-1}$ at 5150~\AA.

\begin{figure}[t]
\plotone{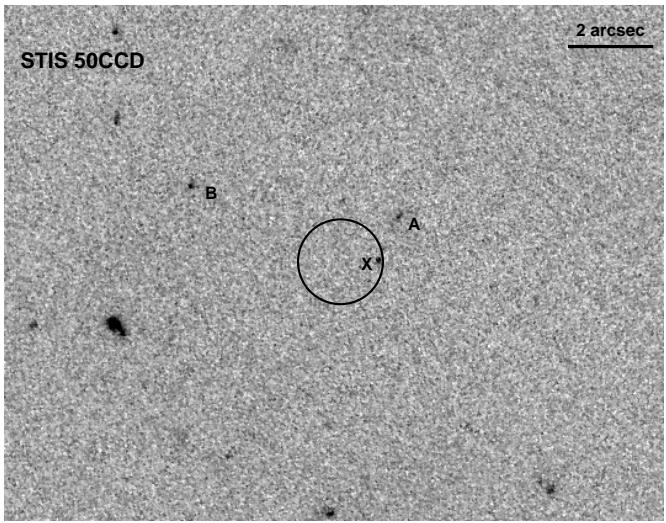}
\caption{\hst/STIS image of the field around \rxj.  The image is
  $\approx 15\arcsec$ on a side, with North up and East to the left.
  The $1\farcs0$-radius \chandra\ error circle is shown.  Source
  X, the likely counterpart of \rxj, and the unrelated sources A and
  B, are also indicated.  Source A is extended.}
\label{fig:stiszoom}
\end{figure}

There is only one optical source inside the \chandra\ error circle.  This
source, marked X in Figure~\ref{fig:stiszoom}, is a possible
counterpart to \rxj.  There are no other potential counterparts
visible in Figure~\ref{fig:stiszoom}, the next closest unresolved source
being $\approx 4\arcsec$ from the \chandra\ position
(source B in Figure~\ref{fig:stiszoom}).

Without color information, it is difficult to accurately photometer
the 50CCD data.  This is because its wide bandwidth makes the aperture
corrections and zero-point fluxes color dependent, leading to
uncertainties of greater than a factor of 2 for the flux coming from
stars ranging from type M to type O.  In what follows, we follow the
analysis of \citet{kkvkf+02} for \rxjk.  We assumed that X is the
counterpart and therefore has blue colors (similar to \rxjw\ and
\rxjk; \citealt{vkk01,kvk98}).  Then we used the bluest of the
available aperture corrections: $0.183$~mag at $0\farcs254$ radius
(T.~Brown 2002, personal communication).  This correction is for a
star with $B-V=-0.09$~mag, compared with an expected $B-V=-0.3$~mag
for \rxj, and is therefore not quite right.  However, the scattered
light that contributes to the color-dependence of the STIS aperture
corrections is predominantly red.  For blue sources, the dependence of
the correction on color is relatively small: for a star with
$B-V=0.05$~mag, the correction changes by about 0.01~mag from that for
a source with $B-V=-0.09$~mag. So the aperture correction used here
should be reasonably appropriate, and to account for any remaining
differences we have added a 0.02~mag systematic uncertainty into the
photometry for \rxj.

With this correction, we find a magnitude of $m=28.56 \pm 0.13$~mag
for X at infinite aperture.  The 3-$\sigma$ limiting magnitude is
$\approx 29.6$~mag.  These magnitudes are in the STMAG system, where
$F_{\lambda} = 10^{-(m+21.1)/2.5}\mbox{ ergs s}^{-1}\mbox{
cm}^{-2}\mbox{ \AA}^{-1}$.  Assuming a spectrum similar to a
Rayleigh-Jeans tail, this relation holds at $\lambda\approx 5148$~\AA\
(this is the wavelength at which a Rayleigh-Jeans spectrum has the
same flux as a flat spectrum that produces the same number of
counts in the 50CCD band; see Appendix~A of \citealt{vkk01}).

From this we find $F_{\lambda}({\rm X})=\expnt{(1.4 \pm
  0.2)}{-20}\mbox{ ergs s}^{-1}\mbox{ cm}^{-2}\mbox{ \AA}^{-1}$ at
  5148~\AA.  We estimate $A_{V}=0.14\pm 0.06$~mag, using the hydrogen
  column from above and the relation from \citet{ps95}.  Again
  assuming a Rayleigh-Jeans spectrum, we convert $A_{V}$ to the
  extinction appropriate for the 50CCD bandpass (again see \citealt{vkk01}) and find $A_{\rm
  50CCD}=0.22\pm 0.09$~mag.  This gives us an unabsorbed flux of
  $\expnt{(1.7\pm 0.3)}{-20}\mbox{ ergs s}^{-1}\mbox{ cm}^{-2}\mbox{
  \AA}^{-1}$.

The optical flux of X is then a factor of $\approx 5$ higher than the
extrapolation of the X-ray blackbody of \rxj, smaller than the value
of 16 found for \rxjw\ \citep{vkk01}, but very similar to the values
found for \rxjk\ and PSR~B0656+14 \citep{kkvkf+02,kpz+01} .  Likewise,
the unabsorbed X-ray-to-optical flux ratio is $\lfxo=4.9$ (where the
X-ray flux has been integrated over the entirety of the blackbody
spectrum).  The similarity of these values to those for other isolated
neutron stars suggests that source X is the optical counterpart of
\rxj.

While a blue color would assure us that X is the counterpart of \rxj,
without color information we cannot be certain.  Source X is very
similar to the counterparts of \rxjk\ and \rxjw, but it is possible
that it is an unrelated source and that no counterpart was detected.
If that is the case, then any counterpart would have $m_{\rm 50CCD} >
29.6$~mag ($\lfxo>5.3$), or an optical flux just consistent with the
extrapolation of the X-ray black-body fit.

Aside from color information (difficult to obtain given its
faintness), another good test for the nature of source X is proper
motion.  Neutron stars have significantly higher proper motions than
the stellar population, with velocities of $\sim 100\mbox{ km s}^{-1}$
typical for the general population of neutron stars \citep*{acc02}.
Such high velocities have been found for the local neutron star
population as well \citep[e.g.,][]{mdlc00,wal01}.  Assuming a velocity
of $100\mbox{ km s}^{-1}$, the proper motion of \rxj\ would be $30
d_{700}^{-1}\mbox{ mas yr}^{-1}$.  While the absolute astrometry from
the STIS image does not have this precision, we expect to be able to
perform relative astrometry with at least $\sim 20$~mas precision (the
limiting factors are distortion correction and modeling of the
point-spread-function, which is color-dependent), although this has
not been tested for STIS.  If this is the case, then in the next few
years proper motion of source X may be detectable, and if so source X
would almost certainly be a neutron star (if X were instead a star, it
would have to be many kpc away and would therefore have negligible
proper motion and be out of the galaxy, given its galactic latitude of
$b=83\degr$).

In the $P$-$\dot P$ plane, \rxj\ appears very similar to the Anomalous
X-ray Pulsars (AXPs; \citealt{m99}).  However, whether or not we have
detected the counterpart of \rxj, the X-ray-to-optical flux ratio is
considerably higher than those found for AXPs
(\citealt*{hvkk00}; \citealt{htvk+01,wc02}): for 4U~0142+61, $\lfxo\approx 4.1$ (where
the X-ray flux is measured from 0.5--10~keV; \citealt{jmcs02}).  The
optical emission from AXPs, which has a non-thermal spectrum, is
thought to arise from the magnetosphere.  Therefore the lack of an
active magnetosphere would significantly decrease the optical flux.
Scaling the non-thermal X-ray emission of 4U~0142+61 by the optical
flux of \rxj, we would predict an X-ray power-law for \rxj\ that would
have been easily visible with \chandra\ ($\expnt{2}{-3}\mbox{ photons
s}^{-1}\mbox{ cm}^{-2}\mbox{ keV}^{-1}$ at 1~keV).  As this power-law
is not seen \citep{hhss02}, it appears that despite its rapid
spin-down \rxj\ does not have an active magnetosphere.  Without an
active magnetosphere, the optical emission from \rxj\ would likely be
similar to those of \rxjw\ and \rxjk, suggesting that we have indeed
found the counterpart to \rxj.

\acknowledgements 

We thank A.~Mahabal for obtaining the LFC data.  D.~L.~K. is supported
by the Fannie and John Hertz Foundation, S.~R.~K by NSF and NASA, and
M.H.v.K. by the Royal Netherlands Academy of Arts and Sciences
KNAW. Based on observations made with the NASA/ESA Hubble Space
Telescope.  Data presented herein were also obtained at the W.~M.~Keck
Observatory, which is operated as a scientific partnership among the
California Institute of Technology, the University of California, and
the National Aeronautics and Space Administration.  The Guide Star
Catalog-II is a joint project of the Space Telescope Science Institute
and the Osservatorio Astronomico di Torino.  The National Radio
Astronomy Observatory is a facility of the National Science Foundation
operated under cooperative agreement by Associated Universities, Inc.

\bibliographystyle{apj}
%\bibliography{magrefs,xray,casA,psrrefs,myrefs}

%% --------------------------------------------------------------------
%% Sat Sep  7 16:27:49 2002
%%   This file was generated automagically from the files
%%   ms.bbl and ms.tex using
%%     /home/quixote/dlk/perl/nat2jour.pl
%%   This file should accompany ms-aas.tex.
%% --------------------------------------------------------------------

\end{document}